# Non-Adlerian phase slip and non stationary synchronization of spin-torque oscillators to a microwave source

G. Finocchio[1], M. Carpentieri[2], A. Giordano,[1] B. Azzerboni[1]

[1] Department of Fisica della Materia e Ingegneria Elettronica, University of Messina, C.da di Dio, I-98100, Messina, Italy.

[2] Department of Elettronica, Informatica e Sistemistica, University of Calabria, Via P. Bucci 42C, I-87036, Rende (CS), Italy.

**Abstract**: The non-autonomous dynamics of spin-torque oscillators in presence of both microwave current and field at the same frequency can exhibit complex non-isochronous effects. A non-stationary mode hopping between quasi-periodic mode (frequency pulling) and periodic mode (phase locking), and a deterministic phase slip characterized by an oscillatory synchronization transient (non-Adlerian phase slip) after the phase jump of $\pm 2\pi$ have been predicted. In the latter effect, a wavelet based analysis reveals that in the positive and negative phase jump the synchronization transient occurs at the frequency of the higher and lower sideband frequency respectively. The non-Adlerian phase slip effect, even if discovered in STOs, is a general property of non-autonomous behavior valid to any non-isochronous auto-oscillator in regime of moderate and large force locking.

I. INTRODUCTION

In the last decade, a class of non-linear auto-oscillators, spin-torque-oscillators (STOs),[1] has been studied experimentally[2,3,4,5] and theoretically[6,7,8] extensively. The STO is promising from a technological point of view being one of the smallest auto-oscillator observed in nature. It also exhibits complex non-linear dynamics arising from the intrinsic coupling between the oscillator phase and power (the effective magnetic field depends on the spatial distribution of the magnetization).[7]

The main properties of STOs are frequency tunability on bias current and field, narrow linewidth, and large output power. In addition, the non-autonomous dynamical behavior of STOs (in presence of microwave current or field) can exhibit non-linear frequency amplitude modulation, frequency pulling, frequency locking, hysteretic and fractional synchronization and stochastic resonance.[9,10,11,12,13,14] Analytical, semi-analytical, and micromagnetic simulations have been used for the prediction or the explanation of those results, which are mainly in regime of "weak" microwave signal, where the oscillator behavior is characterized by isochronous dynamical response and it is possible to neglect the difference between the instantaneous and the stationary (no microwave signal) oscillation power.[7,15,16,17] In the non-isochronous regime ("moderate" or "large" microwave signal), analytical theories fail and a complete numerical approach is necessary, synchronization regions are not symmetric and can be overlapped, and strong non-stationary time domain behaviors (transitions to chaos through period doublings of the orbit, unstable intermittent transition from synchronization to chaos, phase slip) are achieved.[18]

Here, we studied the non-isochronous dynamical behavior of STOs in presence of microwave signal composed by the simultaneous application of microwave current density $J_{AC}$ and field $h_{AC}$, both at the same frequency.

The key result of this paper is the identification of two new dynamical effects in STOs: (i) non-stationary hopping between quasi-stationary Q (frequency pulling) and periodic P (phase locking) mode[19] (the power spectrum is characterized by two modes with power of the same order, one at the frequency of the microwave source and one near the frequency of the self-oscillation mode), and (ii) phase slip characterized by an oscillatory resynchronization transient after a ±2$\pi$ phase jump of the oscillator phase (the power spectrum is characterized by one mode at the frequency of the microwave source and two sidebands). In analogy to the results and the formalism presented in [20], we called this latter effect non-Adlerian phase slip. A wavelet-based time-frequency study shows that when the phase slip occurs, the sideband modes are non-stationary being the high and the low sideband frequency related to the oscillator phase jump of +2$\pi$ and −2$\pi$ respectively.

The paper is organized as follow, Section II introduces the details of the device studied and the numerical implementation of the model, Sections III and IV describes the results and the conclusions of our study.

## II. NUMERICAL DETAILS

We studied the dynamical behavior of exchange biased spin-valves composed by IrMn(8nm)/Py(10nm) (polarizer)/Cu(10nm)/Py(4nm) (free layer) with elliptical cross sectional area (120nm x 60nm) (see inset of Fig. 1(a)). A Cartesian coordinate system has been introduced, where the $x$ and the $y$-axis are related to the easy and the hard in-plane axis of the ellipse respectively. Our numerical experiment is based on the numerical solution of the Landau-Lifshitz-Gilbert-Slonczweski (LLGS) equation.[1] In addition to the standard effective field (external, exchange, self-magnetostatic), the Oersted field and the magnetostatic coupling with the polarizer are taken into account. The time step used was 32 fs. For a complete description of the numerical techniques see also[21]. Typical parameters for the Py have been used: saturation magnetization $M_S$=650x10$^3$ A/m, exchange constant $A$=1.3x10$^{-11}$ J/m, damping parameter $\alpha$=0.02, and polarization factor $\eta$=0.3. The free layer has been discretized in computational cells of 5 x 5 x 4 nm$^3$ (the exchange length is $l_{ex} = \sqrt{\frac{2A}{\mu_0 M_S^2}} \approx 7$ nm). The bias field is applied out-of-plane ($z$-direction) with a tilted angle of 10$^o$ along the $x$-axis to control the in-plane component of the magnetization (see Fig.1(a)).

The polarizer is considered fixed along the *x*-direction. To study the locking, we consider a microwave current $J_{AC} = J_M \sin(2\pi f_{AC} t + \pi/2)$ ($J_M \leq 2\times10^7$ A/cm$^2$) and a microwave field linearly polarized at $\pi/4$ in the *x-y* plane $h_{AC} = h_M \sin(2\pi f_{AC} t + \pi/4)\hat{x} + h_M \sin(2\pi f_{AC} t + \pi/4)\hat{y}$ ($h_M \leq 3$ mT). This microwave field can be generated by using the experimental technique developed in ref[12]. The computational data presented in the rest of the paper have been performed with no thermal effects.

## III. RESULTS AND DISCUSSIONS

(A) *Free running data*

First of all, we characterized the STO in the free running regime (no microwave signal). Persistent magnetization oscillation is observed in a wide range of current density and for out-of-plane bias field larger than 180 mT. Here we discuss in detail data for a bias field of 200 mT, but qualitative similar results have been also achieved for 180 and 220 mT. Figs. 1(a) and (b) display current density *J* dependence of the oscillation frequency ($f_0$) and the integrated output power for the Giant Magneto-Resistive (GMR) signal (*H*=200 mT). The $f_0$ vs *J* curve is characterized by red shift and an in-plane oscillation axis up to $J_1$=-3.4x10$^7$ A/cm$^2$ while for |*J*|>|$J_1$| the magnetization precesses around an out-of-plane axis (blue shift). The discontinuities observed in the data are related to oscillation axis jumps.[22, 24]

(B) *Isochronous Synchronization*

We systematically studied the locking region to the first harmonic (the same of the self-oscillation) as function of $J_M$ and $h_M$ in the blue shift region. Fig. 1(c) shows the locking region computed for *J*=-5 and -8x10$^7$ A/cm$^2$ and related to a microwave current only ($h_M$ =0 mT). The response can be considered in the regime of "weak" microwave signal, in fact the locking regions are symmetric and the locking band is linearly dependent on $J_M$. As already predicted,[23] it is also found an intrinsic phase shift $\Psi_I$ between the phase of the self-oscillation $\Psi$ and the phase of the microwave current $\Psi_E$ in the whole locking region. Fig. 1(d) summarizes $\Psi_I$ as function of $f_{AC}$, as can be observed a linear relationship is achieved and, depending on $f_{AC}$ and $J_M$, $\Psi_I$ can also assume the value 0 or $\pi/2$.

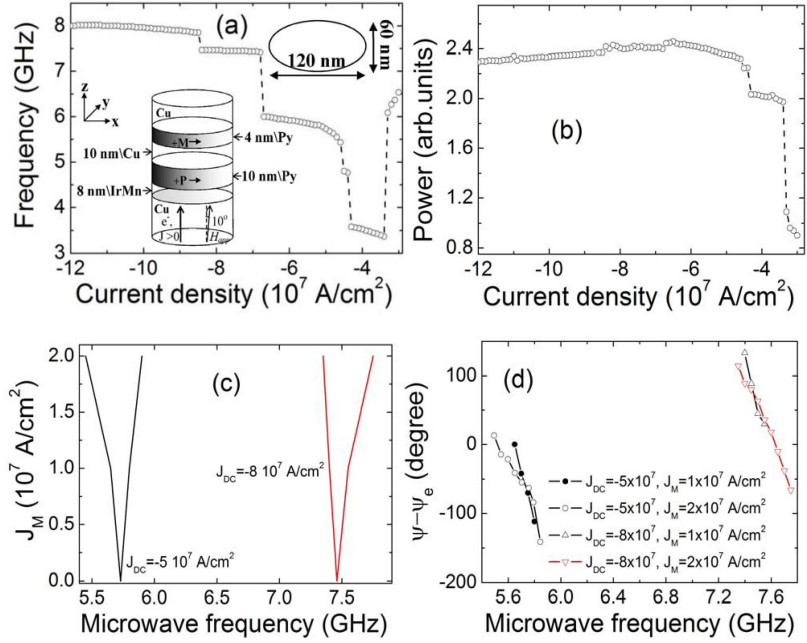

FIG. 1 (a) Auto-oscillation frequency $f_0$ as function of the current density ($H$=200 mT) (inset: sketch of the studied device); (b) integrated output power as function of the current density; (c) locking regions computed for $J$= -5 and -8x10$^7$ A/cm$^2$ (only a microwave current is applied); (d) intrinsic phase shift $\Psi_I$ between the phase of self-oscillation and the microwave current in the locking region of Fig. 1(c) (both $J$=-5 and -8x10$^7$ A/cm$^2$) for $J_M$=1 and 2x10$^7$ A/cm$^2$.

(C) *Non-Isochronous Synchronization*

As will be discussed below, when both microwave current and field are applied simultaneously at the same frequency, the non-autonomous response becomes more complicated and non-isochronous effects are observed in the locking region. Fig. 2(a) shows the dependence of the frequency of the excited mode with the larger power as function of the microwave frequency for $J$=-5x10$^7$ A/cm$^2$, $J_M$ =1x10$^7$ A/cm$^2$, and $h_M$ =1 mT, it can be observed that the presence of the microwave field gives rise to an increasing of the locking region (from 150 MHz at $h_M$ =0 mT up to 1.3 GHz for $h_M$ =1 mT) and to a behavior different from the one described in Figs.1(c) and (d). We performed a systematic study to better understand the origin of this dynamical behavior and Fig. 2(b) summarizes the locking region ($J$=-5x10$^7$ A/cm$^2$) computed up to $J_M = 1 \times 10^7$ A/cm$^2$ ($h_M = 0$ mT) and then increasing $h_M$ up to 3 mT maintaining fixes $J_M = 1 \times 10^7$ A/cm$^2$. The border lines have been computed considering the last microwave frequency where a mode at the frequency of the microwave source with large power is excited. As can be observed, this locking region is asymmetric and at some microwave frequencies the power spectrum cannot be identified as a regular P-mode (see Fig. 2(c)). The results obtained for $J=-5 \times 10^7$ A/cm$^2$ ($J_M = 1 \times 10^7$ A/cm$^2$ and $h_M = 1$ mT) are discussed in detail. Fig. 2(c) shows the power spectra related to the points 1-3 depicted in Fig. 2(b), the spectrum for $f_{AC}$=5.75 GHz (point 1) is a regular P-mode, the spectra at 5.95 GHz (point 3) and 5.2 (point 2) are the evidence of the strong non isochronisms, in fact together with the mode with the large power at the microwave frequency (locking regime), an additional large power mode or two sidebands are excited respectively. Qualitative similar results have been also obtained up to $J=-9.5 \times 10^7$ A/cm$^2$ and for $h_M$ =2-3 mT and $J_M$ =1.5 and $2 \times 10^7$ A/cm$^2$.

The origin of this non-isochronisms can be understood by means of a time-frequency study of the GMR-signal based on the wavelet transform. We used the complex Morlet as mother wavelet:

$$\psi_{u,s} = \frac{1}{\sqrt{s\pi f_B}} e^{j2\pi f_c \left(\frac{t-u}{s}\right)} e^{-\left(\frac{t-u}{s}\right)^2 / f_B} \qquad (1)$$

where $f_C$ and $f_B$ are two parameters which characterize the mother wavelet function, $s$ and $u$ are the scale and the shift parameters respectively. The wavelet transform $W_r(u,s)$ is computed as:

$$W_r(u,s) = \frac{1}{\sqrt{s}} \int_{-\infty}^{+\infty} r(t) \psi^* \left(\frac{t-u}{s}\right) dt \qquad (2)$$

where $r(t)$ is the time domain GMR signal and the $\psi^*$ is the complex conjugate of the mother wavelet (see[24] for all the computational details). Fig. 2(d) displays the wavelet scalogram (the modulus of $W_r(u,s)$) for the time trace related to the point 3 in Fig. 2(b) with $f_C$=1 and $f_B$=300 (the amplitude increases from black to white). The wavelet analysis shows a mode hopping between the two modes, in detail, in some time ranges (for example 5.5-9 ns) the P-mode (phase locking) is excited while in other time ranges (for example 9.5-11.5 ns) is the Q-mode (frequency pulling) to be excited (the main mode has the frequency near the self-oscillation mode and a low power peak at the microwave frequency). We refer to this behavior as non-stationary P/Q mode. The relative power of the two modes depends on how long they are excited, for example at $f_{AC}$ =5.95 GHz the power of the P-mode is larger than the one of the Q-mode.

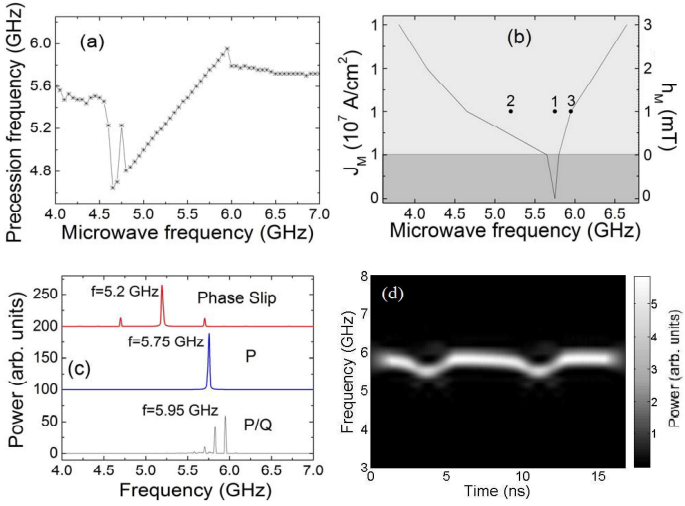

FIG. 2: (a) Frequency of the main excited mode (the mode with the largest power) as function of the microwave frequency computed for $J$=-5x10$^7$ A/cm$^2$, $J_M$ =1x10$^7$ A/cm$^2$ and $h_M$ =1 mT; (b) Locking region computed for $J$=-5x10$^7$ A/cm$^2$; (c) power spectra for P-mode ($f_{AC}$=5.75 GHz, point 1), P/Q non-stationary mode ($f_{AC}$=5.95 GHz, point 3), and non-Alderian phase slip ($f_{AC}$=5.2 GHz, point 2), $J_M$ =1x10$^7$ A/cm$^2$ and $h_M$ =1 mT as indicated in (b); (d) wavelet transform of the time trace of the GMR signal related to the P/Q mode of (c).

In several non-linear systems a diffuse non-isochronous effect called phase slip is observed.[25, 26] It is characterized by brief periods of resynchronization after a phase jump of $\pm 2\pi$ of the oscillator phase. Commonly, after the phase jump the resynchronization transient occurs asymptotically (Adlerian phase slip).[23] Differently, here we observe in the locking region the P-mode coupled with two sidebands. The wavelet transform of those time domain GMR signals displayed in Fig. 3(a) for $f_{AC}$ =5.2 GHz (point 2 of Fig. 2(b)) indicates the non-stationary excitation of the sideband modes, result which is completely different from the modulation processes where the sideband modes are stationary.[9, 15] In particular, the time evolution of the oscillation phase Ψ is characterized by the occurrence of a phase slip with an oscillatory synchronization transient we called non-Adlerian phase slip (Fig. 3(b) summarizes those computations for 2 ns). The origin of the non-stationary sidebands is related to the fact that the higher and lower sideband modes are connected to the resynchronization frequency of a $+2\pi$ or $-2\pi$ phase slip.

From a theoretical point of view, the linewidth of a regular P-mode coincides with the one of the microwave source, the phase slip introduces an additional intrinsic dissipation mechanism that can be seen as enhancement of the oscillator linewidth and should be taken into account in the design of arrays of STOs which should work synchronized. The presence of the phase slip can also explain for example the experimental evidence of the linewidth enhancement in the power spectrum of STOs near the boundary of the locking region (see Fig. 3(d) of Ref.[13]) and the thermally induced sideband observed in Ref.[27].

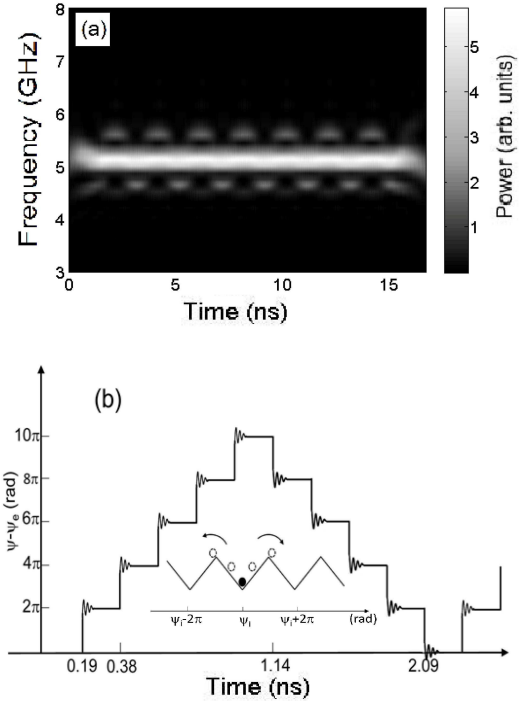

FIG. 3: (a) Wavelet scalogram (white/black color corresponds to the largest/smallest wavelet coefficient amplitude) for $J=-5\times10^7$ A/cm$^2$ and $J_M=1\times10^7$ A/cm$^2$ and $h_M=1$ mT at the frequency of $f=5.2$ GHz. (b) Time domain evolution of the phase of the oscillator. Inset: example of energy landscape where it is possible to observe positive or negative phase jump.

We point out the fact that the phase slip achieved here being periodic (deterministic phase slip) at the same frequency of the microwave source is different from the one obtained in a non-linear system where the thermal fluctuations are responsible of the phase jumps among different energetic minima (stochastic phase slip). To understand the origin of the observed deterministic phase slip, we introduce the simple scenario displayed in the inset of Fig. 3(b), when the oscillation induced by the microwave source is large enough to overcome the energy barrier which separates two different minima (this is achieved for some value of $f_{AC}$) the phase slip occurs.

In addition, our computations suggest that the discontinuities in the free running data reduce the amplitude of the microwave source to apply in order to excite the phase slip and the non-stationary P/Q mode. In fact, the synchronization data (by using the same microwave signals) achieved for an out of plane bias field larger than 400 mT, where discontinuities in the free running data disappear, are similar to the results already published in literature and described analytically in the regime of "weak" microwave signal (see for example Ref.[7] for a review).

IV. SUMMARY AND CONCLUSIONS

Fig. 4 summarizes qualitatively the time dependence of the oscillator phase for the Q-mode (frequency pulling) where $\Psi_I$ increases linearly, the P-mode (phase locking) where $\Psi_I$ is constant, the non-stationary P/Q region where exists a time domain mode hopping between Q and P mode (the phase can be constant or increases linearly), and the Adlerian and non-Adlerian phase slip (abrupt jumps of $2\pi$ in the oscillator phase are observed together to asymptotic and oscillatory resynchronization).

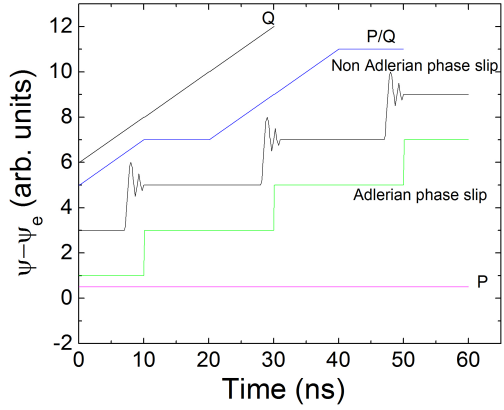

FIG. 4: Time evolution of the oscillator phase for different excited modes: Q-mode, non-stationary P/Q mode, non-Adlerian and Adlerian phase slip, and P-mode.

In summary, in the non-autonomous behavior of STOs by means of micromagnetic simulations and time-frequency domain analysis based on wavelet transform, the presence of the non stationary P/Q modes and non-Adlerian phase slip is found. The key finding is the possibility to drive deterministic phase slip by means of the application of a moderate microwave source that can be easily generated experimentally. We think our results can stimulate future experimental studies of the non-isochronisms in self-oscillators such as non-stationary behavior and phase slip not thermally activated. Finally, we identified a phase slip with a an oscillatory resynchronization transient, non-Adlerian phase slip, that even if discovered in STOs, is a general property of non-autonomous behavior valid to any auto-oscillator in regime of moderate and large amplitude microwave signal.

## ACKNOWLEDGES

This work was supported by Spanish Project under Contract No. MAT2011-28532-C03-01. The authors would like to thank Prof. Sergio Greco for his support with this research.

[19] In the regime of isochronisms, a STO with $f_0$ as stationary oscillation frequency can exhibit different behavior depending on the frequency $f_{AC}$ of the microwave signal such as frequency pulling and phase locking. The frequency pulling occurs when the $f_{AC}$ approaches $f_0$ and a quasi-stationary mode or Q-mode is excited (the power spectrum is characterized by two modes one near $f_0$ and one with small power at $f_{AC}$). The phase locking occurs when the oscillation frequency $f_0$ is locked at the frequency of the

external microwave source $f_{AC}$ and a periodic mode or P-mode is excited (the power spectrum is characterized by a single mode at $f_{AC}$).